# Predicting Fine-grained Behavioral and Psychological Symptoms of Dementia Based on Machine Learning and Smart Wearable Devices

Benny Wei-Yun Hsu, Yu-Ming Chen, Yuan-Han Yang, Vincent S. Tseng

***Abstract*—Behavioral and Psychological Symptoms of Dementia (BPSD) impact dementia care substantially, affecting both patients and caregivers. Effective management and early detection of BPSD are crucial to reduce the stress and burden on caregivers and healthcare systems. Despite the advancements in machine learning for dementia prediction, there is a considerable gap in utilizing these methods for BPSD prediction. This study aims to fill this gap by presenting a novel personalized framework for BPSD prediction, utilizing physiological signals from smart wearable devices. Our personalized fine-grained BPSD prediction method accurately predicts BPSD occurrences by extracting individual behavioral patterns, while the generalized models identify diverse patterns and differentiate between various BPSD symptoms. Detailed comparisons between the proposed personalized method and conventional generalized methods reveals substantial improvements across all performance metrics, including a 16.0% increase in AUC. These results demonstrate the potential of our proposed method in advancing dementia care by enabling proactive interventions and improving patient outcomes in real-world scenarios. To the best of our knowledge, this is the first study that leverages physiological signals from smart wearable devices to predict BPSD, marking a significant stride in dementia care research.**

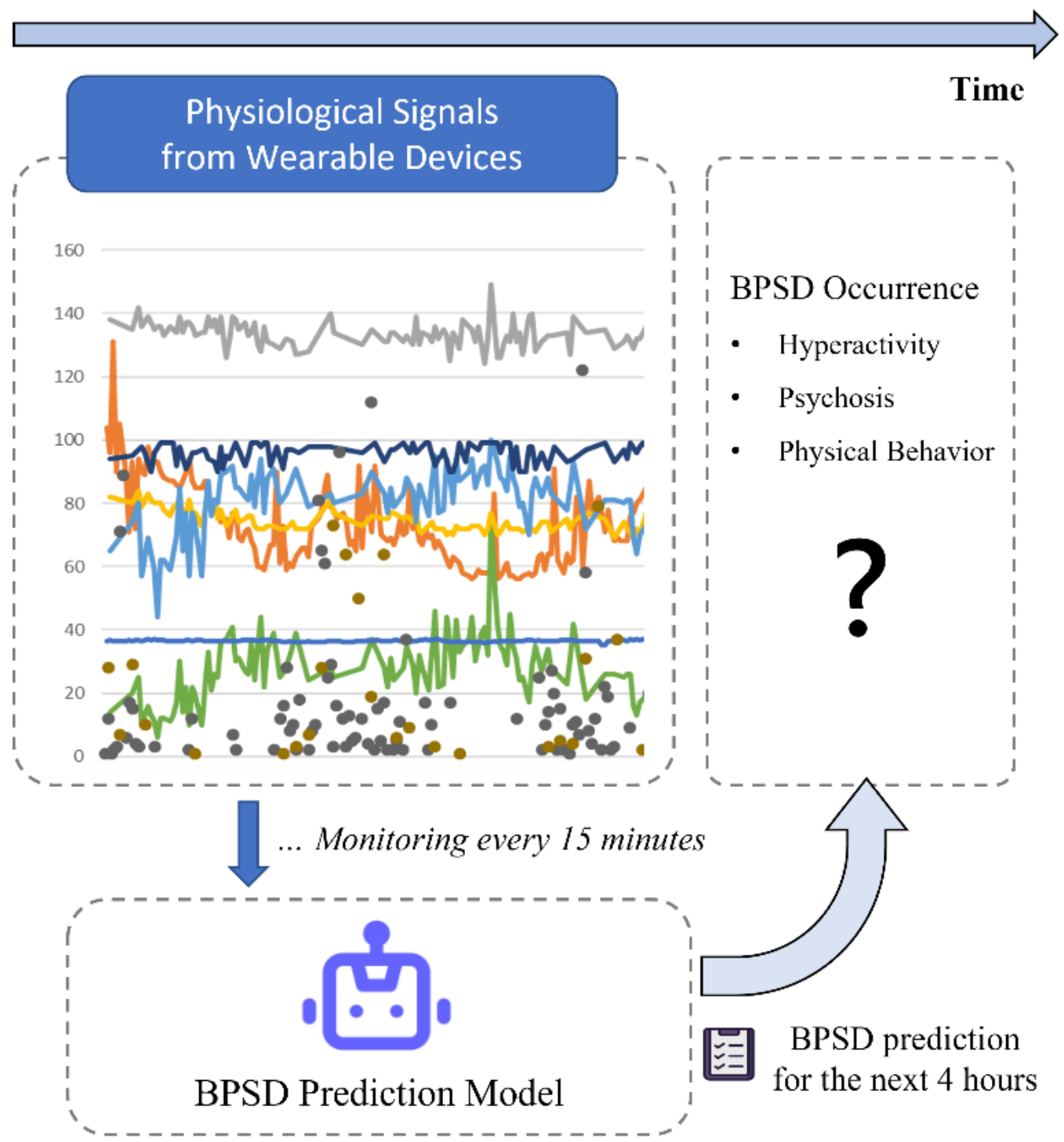

Fig. 1. Overview of the proposed workflow. The dementia patient's physiological signals captured from smart wearable devices were fed into the machine learning model to predict the occurrences and type of BPSD.

## I. INTRODUCTION

BEHAVIORAL and Psychological Symptoms of Dementia (BPSD) represent a diverse range of non-cognitive symptoms commonly manifest in individuals with dementia. These symptoms include disinhibition, anxiety, elation/euphoria, agitation/aggression, hallucinations, delusions, apathy/indifference, irritability, appetite and eating abnormalities, aberrant motor behavior, depression/dysphoria, and sleep and night-time behavior disturbances occurring at any stage of the disease that constitutes a vital clinical aspect of dementia [1][2]. The management of BPSD poses significant challenges for clinical applications and caregivers, as it not only impacts the well-being and quality of patients' lives but also leads

Benny Wei-Yun Hsu is with the Institute of Computer Science and Engineering, National Yang Ming Chiao Tung University, Hsinchu City 300093, Taiwan R.O.C. (email: bennyhsu520@gmail.com).

Yu-Ming Chen is with the Department of Computer Science, National Yang Ming Chiao Tung University, Hsinchu 300093, Taiwan, R.O.C. (email: tim310579.cs06@nctu.edu.tw)

Ting-Yi Lin is with the Neuroscience Research Center, Kaohsiung Medical, University, Kaohsiung, Taiwan, and also with the Department of Neurology, Kaohsiung Municipal Ta-Tung Hospital, Kaohsiung, Taiwan, R.O.C. (email: tilin201909@gmail.com).

Yuan-Han Yang is with the Neuroscience Research Center, Kaohsiung Medical, University, Kaohsiung, Taiwan, and also with the Department of Neurology, Kaohsiung Municipal Ta-Tung Hospital, Kaohsiung, Taiwan, R.O.C. (email: endlessyhy@gmail.com).

Vincent S. Tseng is with the Department of Computer Science, National Yang Ming Chiao Tung University, Hsinchu 300093, Taiwan, R.O.C. (email: vtseng@cs.nycu.edu.tw).

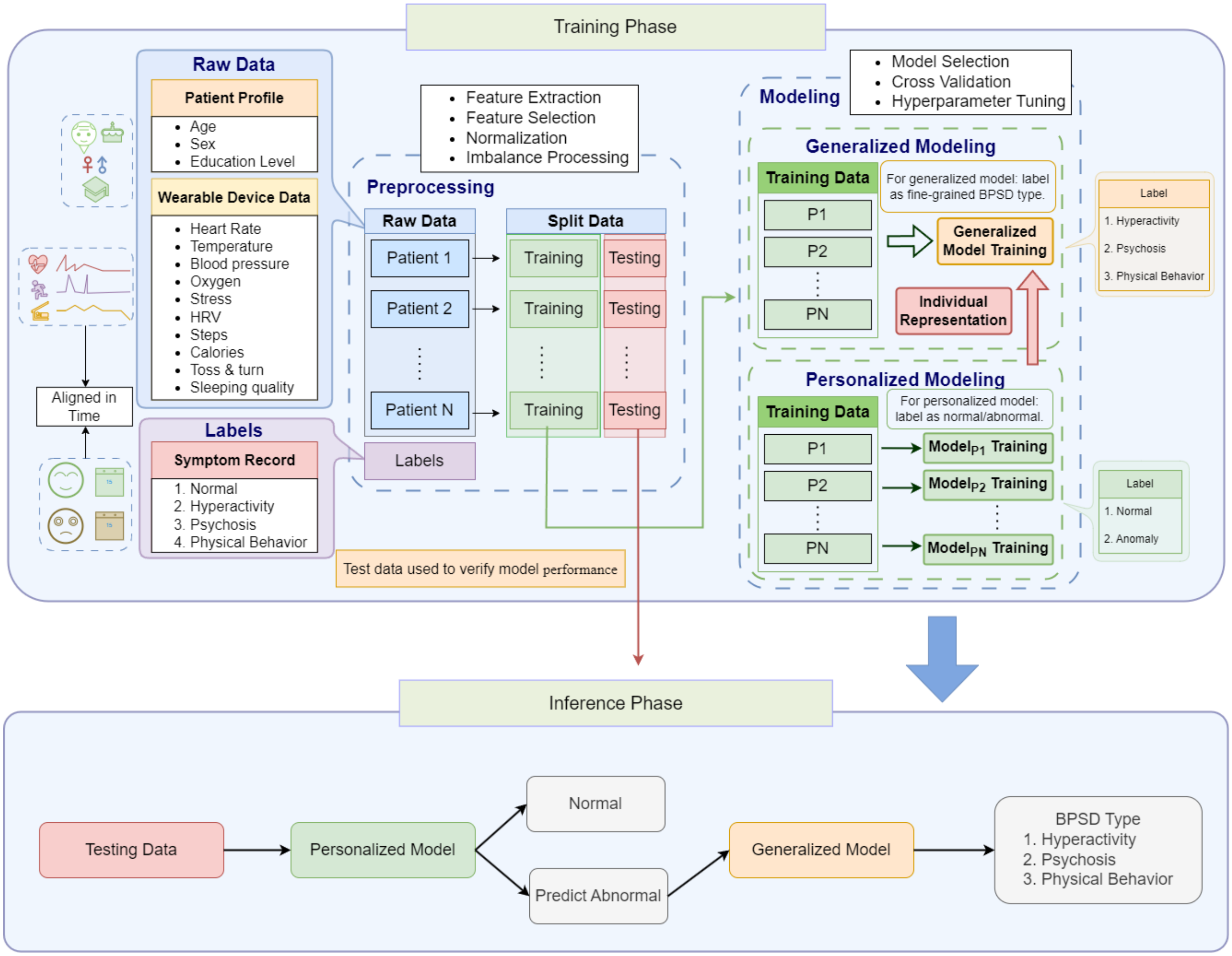

Fig. 2. The proposed framework. In the training phase, the data consists of basic information (age, sex, and education level) and physiological signals from the wearable device, which are aligned with the time points of BPSD records. After data preprocessing, it proceeds to the modeling phase to generate general and personalized models, for fine-grained BPSD prediction and BPSD occurrence prediction, respectively. In the inference phase, the data is passed to the personalized model to find whether the abnormal symptoms will occur in the future. If labeled as abnormal, the data will go through the generalized model to determine the fine-grained BPSD type.

to elder abuse, caregiver stress, and accelerated functional decline in individuals with dementia [3][4][5][6]. Additionally, the presence of BPSD is associated with several adverse outcomes, including increased hospitalization risks, impaired daily functioning, diminished quality of life, and a higher likelihood of progressing to severe dementia at a faster rate [7][8][9]. Consequently, the early detection and effective management of BPSD assume critical importance in providing timely interventions, support, and care that can positively influence patient outcomes and alleviate the burden on caregivers and healthcare resources. Nevertheless, there is a limited number of studies related to BPSD prediction based on machine learning methods. Cho et al. [12] presented a machine learning-based approach for the next-day BPSD prediction. However, with such a large time frame for predicting whether BPSD will occur the next day, it is not effectively aiding caregivers in early intervention for real-world applications.

To address the aforementioned problems, this study proposed a BPSD prediction framework involving personalized and generalized models based on temporal data from smart wearable devices. Personalized models utilize the patient's unique information to extract individual behavioral patterns from wearable devices for predicting the occurrence of BPSD. On the contrary, generalized modeling methods were designed to capture patterns and trends across a diverse population, aiming to develop a predictive framework that applies broadly to different individuals.

TABLE I
DESCRIPTION OF THE VARIABLES USED FOR THIS STUDY.

| Variable | Description | Measurement occasion |
|---|---|---|
| Age | Patient age | - |
| Sex | Patient sex | - |
| Education Level | Education level (year) of patient | - |
| HR | Heart rate | Continuous[1] |
| HRV | Heart rate variability | Continuous |
| SYS | Systolic blood pressure | Continuous |
| DIA | Diastolic blood pressure | Continuous |
| Stress | Stress index | Continuous |
| Temp | Body temperature | Continuous |
| Oxygen | Blood oxygen | Continuous |
| Steps | Steps | Activity[2] |
| Calories | Calories burned | Activity |
| Toss & turn | Number of tosses & turns | Sleep[3] |
| Sleeping quality | Sleeping quality | Sleep |

1. Passive measurement while wearing.
2. Measurement during activity.
3. Measurement during sleeping.

If the personalized model predicted a patient abnormality, the generalized model would be used to classify BPSD symptoms, including hyperactivity, psychosis, and physical behavior [1]. Additionally, the proposed framework predicts BPSD occurrences and symptoms in intervals of fifteen minutes over a four-hour period, enabling caregivers to intervene proactively for dementia patients.

An overview of the workflow is depicted in Fig. 1. This study has contributed to advancing dementia care practices for improving patient outcomes in geriatric medicine by proposing the BPSD prediction framework. The prominent contributions are summarized as follows:

- We have developed the first personalized framework to predict BPSD occurred in the next four hours based on physiological signals from smart wearable devices. This innovation empowers efficient and timely interventions in dementia care.
- We have conducted a series of experiments to validate the effectiveness of our proposed fine-grained BPSD prediction method, outperforming the conventional machine learning approaches.
- We analyze the novel personalized fine-grained BPSD prediction models, indicating promising research directions for techniques development in the dementia care field.

## II. RELATED WORKS

In recent years, the application of advanced machine learning and data analytics techniques in the medical domain has gained significant attention [10]. Dementia research, as a critical area within the medical field, has seen a surge in studies employing machine learning methods [11]. These studies aim to predict the onset of dementia using a diverse range of input data types, including magnetic resonance imaging (MRI) [12][13], positron emission tomography (PET) scans [14], cognitive test data [15], genetic information [16], and medical records [17]. In Alzheimer's disease (AD) research, neuroimaging data from MRI and PET scans have proven effective in AD classification. Cognitive test results, such as cognitive function scales, have also been widely utilized to forecast dementia onset. Moreover, genetic data has been explored extensively to understand its associations with dementia. However, when addressing the prediction of BPSD, the need arises for easily accessible data that can substantially support clinical caregiving.

There are limited studies regarding BPSD prediction with machine learning approaches. Cho et al. [12] presented the first study using demographic, cognitive, and functional status, actigraphy, personality type, and symptom diary data for BPSD prediction. It is also the only study to consider adopting an accelerometer for BPSD prediction research. However, their study focused on the next-day prediction, a wide range of predictive time windows, making it hard to target the particular period for early intervention. It also needed several types of data that require more cost for each patient. Therefore, although their study proposed BPSD prediction models, a gap still existed for real-world applications.

## III. PROPOSED METHODS

### A. Data Description

The initial study cohort comprised 250 cases. Subsequently, cases with records spanning less than two weeks were excluded. This selection process yielded a dataset of 183 confirmed dementia patients, consisting of 101 females and 82 males. The average age is 77.9 ± 10.2 (female: 77.7 ± 8.8; male: 78.2 ± 11.7), the average education level (year) is 6.6 ± 4.5years (female: 5.0 ± 4.2; male: 8.6 ± 4.1). In addition, caregivers observe the daily behaviors of the patients and record the abnormal behaviors or symptoms related to dementia that occur in their daily lives. These records are used as labels for subsequent prediction. The symptom records are first aligned with the values measured by the wearable device in time, and the physiological values measured every 15 minutes were used to predict the type of BPSD symptoms that will occur in the next four hours. Each participant had physiological signal data collected over two to four weeks. The patients wore the wearable smart watches approximately between 8:00 a.m. and 5:00 p.m., and this was conducted around 3-5 days weekly. The data composition is presented in TABLE I. To timely predict BPSD occurrences, we used current physiological values and past records comprising the values of the previous 15, 30, 45, and 60 minutes. In other words, there are 11 physiological signals during measurement at five continuous time points, with a total of 55 physiological values for BPSD prediction. For each patient, their records were segmented into multiple periodic entries after removing the cases without BPSD labels, and each entry was categorized as one of the following categories: Normal (n = 20,529), Hyperactivity (n = 274), Psychosis (n = 2,564), or Physical Behavior (n = 1,394) symptoms.

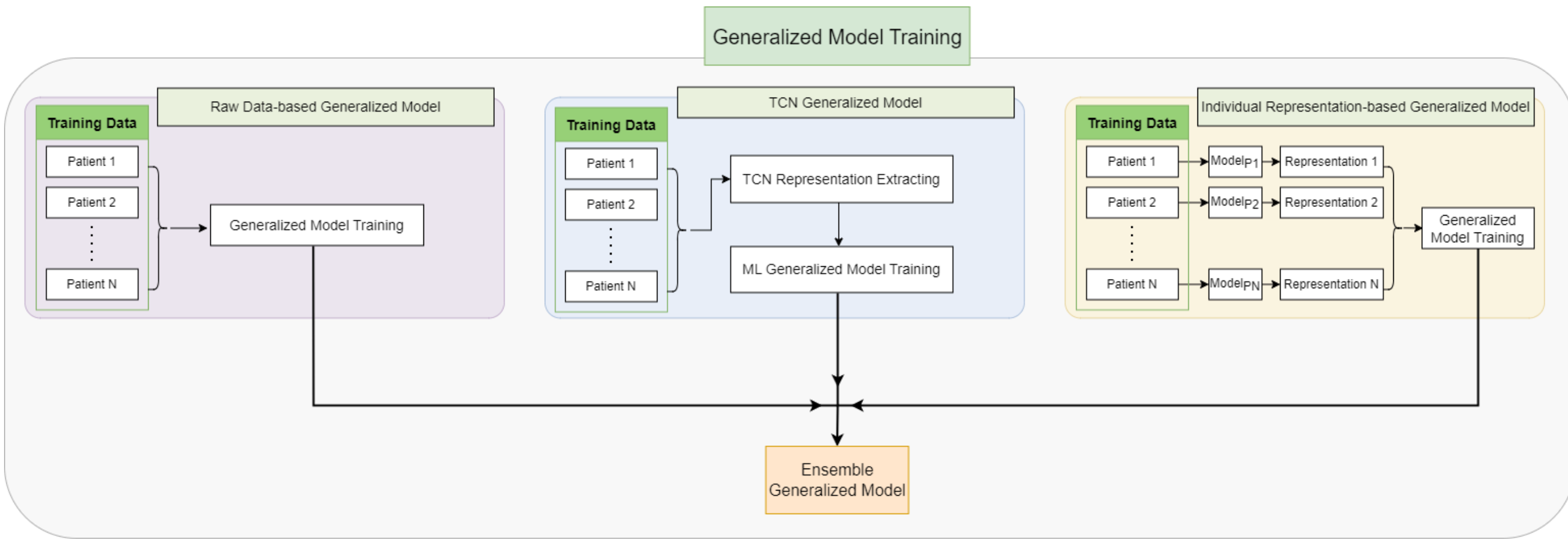

Fig. 3. Workflow of generalized model training. There are three types of generalized models adopted, which are Raw Data-based (RDG), TCN, and Individual Representation-based (IRG) generalized model. First, RDG is used for learning the raw information of the data. Second, TCN is used to capture the temporal dependency in the raw time series data. Third, IRG extract features generated from the personalized model include the specialized individual information of each patient. These generalized models will be incorporated to classify the BPSD.

## B. *Fine-grained BPSD Prediction*

Our proposed framework is illustrated in Fig. 2. The data used in this study consists of physiological signals measured when dementia patients wear smart wearable devices, as well as patients' basic information, i.e., age, sex, and education level. The first step is data preprocessing, containing feature extraction, feature selection, normalization, and imbalance processing. The modeling process consists of personalized and generalized modeling phases. The personalized focused on individual differences, where each patient's data was used to build a unique model, and it was employed to predict the occurrence of BPSD. In contrast, the generalized model was employed to predict the specific type of BPSD. We adopted a two-stage architecture because not all patients exhibit the same type of BPSD, making it impractical to build an individual multiclass BPSD model for each patient. Moreover, the personalized model can accurately capture patients' unique representations, and the generalized model was dedicated to differentiating between different fine-grained BPSD types, leveraging the collective data from all patients. For the personalized model, we employ a separate model for each patient, where we adopted Extremely Randomized Trees (ERT) [19] to construct each personalized model. ERT evolved from Decision Tree [20] and Random Forest [21] algorithms. The main idea behind tree-based methods is to build a hierarchical structure of decision rules based on input features. A decision tree (DT) consists of nodes and branches, where each internal node represents a feature and a decision rule, and each leaf node represents a prediction or class label. Random forest (RF) combines multiple decision trees for prediction. Each tree is trained on a random subset of the training data, and the final prediction is obtained by aggregating the prediction results of all trees, which can handle high-dimensional data and reduce overfitting. ERT is similar to random forests. However, they make more random selections for feature splitting points when constructing each decision tree, thereby increasing model diversity and generalization capability.

The main objective of the personalized models is to predict whether symptoms will occur in the next four hours, which can be represented as follows:

$$L = M\,(F) \tag{1}$$

where F is the patient information and features measured by the wearable devices in the past hour. The M represents the model used for predicting the symptoms. L is the label that the patient will experience BPSD, or nothing happened in the next four hours.

The generalized model consists of three components as depicted in Fig. 3. First, we aggregated all the patients' records to train a generalized model, which is named

TABLE II
PERFORMANCE COMPARISON WITH THE PROPOSED PERSONALIZED AND THE CONVENTIONAL GENERALIZED METHODS.

| Method | AUC | Sensitivity | Specificity | Accuracy | Precision | F1-Score |
|---|---|---|---|---|---|---|
| Conventional Generalized Method | 0.701 | 0.472 | 0.647 | 0.744 | 0.484 | 0.441 |
| **Personalized Method (Ours)** | **0.813 ± 0.129** | **0.741 ± 0.152** | **0.858 ± 0.085** | **0.771 ± 0.1438** | **0.685 ± 0.150** | **0.661 ± 0.153** |

TABLE III
PERFORMANCE COMPARISON OF CONVENTIONAL CLASSIFIERS WITH OUR PREPROCESSING APPROACHES.

| Model | AUC | Sensitivity | Specificity | Accuracy | Precision | F1-score |
|---|---|---|---|---|---|---|
| LR | 0.740 ± 0.145 | 0.615 ± 0.174 | 0.852 ± 0.097 | 0.706 ± 0.176 | 0.582 ± 0.148 | 0.549 ± 0.155 |
| RF | **0.775 ± 0.141** | **0.706 ± 0.159** | 0.833 ± 0.089 | **0.741 ± 0.150** | **0.627 ± 0.141** | **0.607 ± 0.147** |

TABLE IV
PERFORMANCE COMPARISON OF BACKBONE CLASSIFIERS WITH OUR PROPOSED FRAMEWORK.

| Backbone | AUC | Sensitivity | Specificity | Accuracy | Precision | F1-score |
|---|---|---|---|---|---|---|
| LR | 0.793 ± 0.130 | 0.690 ± 0.145 | **0.857 ± 0.092** | 0.733 ± 0.159 | 0.655 ± 0.141 | 0.620 ± 0.142 |
| RF | **0.818 ± 0.120** | **0.739 ± 0.149** | 0.855 ± 0.092 | **0.764 ± 0.149** | **0.675 ± 0.137** | **0.654 ± 0.146** |

Raw-Data-based Generalized model (RDG). We used three machine learning algorithms, including Extremely Randomized Trees (ERT), Random Forest (RF), and Logistic Regression (LR), to assess the performance of the backbone models. Second, we employed a Temporal Convolutional Network (TCN), a widely used and effective method for handling time series data [22], to build a generalized model for temporal feature extraction from the raw data. Subsequently, the latent representation with a 256-dimensional vector in the training phase was extracted, which serves as the input to train a classifier. Third, we adopted an Individual Representation-based Generalized model (IRG), which considers each patient's distinctive feature in the training phase, emphasizing the refined information from personalized models. After training the three generalized models, we employed an ensemble approach that can concurrently consider different aspects for predicting BPSD and enhancing performance. The main objective of the generalized model is to predict the category of symptoms that will occur in the next four hours for each data after the personalized model prediction, which can be represented as follows:

$$C = E\,(G1\,(F),\, G2\,(F),\, G3\,(F)) \tag{2}$$

## IV. EXPERIMENTAL EVALUATION

### A. Experimental Settings

The training data from each patient with BPSD records were aggregated to form a generalized dataset. This dataset, comprising data from all patients, was used to train a generalized model. We employed 5-fold cross-validation and fine-tuned hyperparameters to optimize the performance of this generalized model, which was based on Extra Randomized Trees (ERT). Personalized data for each patient was split into a training set (70%) and a testing set (30%). It is important to note that personalized models were only constructed for patients with records of BPSD occurrences during the measurement period. Patients without such records were excluded from the personalized model building process.

During the testing phase, the split personalized test data was initially processed by the personalized model to detect the occurrence of BPSD. If the personalized model predicted BPSD occurrence, the data was subsequently passed through the generalized model to classify the specific type of BPSD. The entire assessment process was executed independently for each patient. As a result, the framework's performance was evaluated for each patient separately. To provide an overall assessment of the framework's effectiveness in predicting BPSD occurrences and types, we calculated the average performance across all patients. Performance evaluation of the models includes the area under the ROC curve (AUC), sensitivity, specificity, accuracy, precision, and F1-score.

### B. Performance Comparison of Personalized and Generalized Frameworks

To investigate the distinction between the proposed personalized and conventional generalized frameworks, we conducted an experiment utilizing Extra Randomized

TABLE V
OVERALL PERFORMANCE WITH ABLATION STUDIES ON GENERALIZED MODELS.

| Combination | AUC | Sensitivity | Specificity | Accuracy | Precision | F1-Score |
|---|---|---|---|---|---|---|
| RDG | 0.795 ± 0.129 | 0.734 ± 0.156 | 0.842 ± 0.090 | 0.760 ± 0.149 | 0.643 ± 0.141 | 0.626 ± 0.153 |
| RDG + TCN | 0.779 ± 0.143 | 0.559 ± 0.140 | 0.850 ± 0.101 | 0.544 ± 0.199 | 0.645 ± 0.213 | 0.484 ± 0.157 |
| RDG + IRG | 0.808 ± 0.129 | 0.733 ± 0.164 | 0.846 ± 0.090 | 0.762 ± 0.147 | 0.642 ± 0.140 | 0.629 ± 0.150 |
| RDG + IRG + TCN | **0.813 ± 0.129** | **0.741 ± 0.152** | **0.858 ± 0.085** | **0.771 ± 0.144** | **0.685 ± 0.150** | **0.661 ± 0.153** |

Trees (ERT) as the underlying algorithm. In the conventional generalized model, the personalized modeling stage was omitted, and it was directly trained on the four-class dataset, containing labels for normal, hyperactivity, psychosis, and physical behavior. The training and testing distributions followed the configuration illustrated in Fig. 2, with the individual subsets of training and testing sets merged separately (i.e., combining the original training sets of each individual into one training set, and similarly for the testing set). Hence, the conventional generalized model in this experiment generated a single model for evaluation to ensure a fair comparison. TABLE II presents the comparative results. As per the findings, the proposed personalized model consistently outperformed the conventional generalized model across all evaluation metrics. The personalized model exhibited a remarkable 16.0% increase in AUC compared to the generalized model, while sensitivity and specificity significantly improved 57% and 32.6%, respectively. The precision rate and F1-score also exhibited considerable enhancements.

The limitations of the conventional generalized model in accurately predicting fine-grained BPSD based on wearable device data become prominent. The reason lies in the substantial variation in normal behaviors among dementia patients. The conventional framework was developed to fetch common patterns to distinguish normal and abnormal behaviors, but the behavioral baseline varies significantly among dementia patients. Consequently, the proposed personalized framework sets a new standard for research in smart wearable device-based BPSD prediction.

### C. Analysis of Personalized Framework with Various Backbone Models

TABLE III presents the results of our experiments, showcasing the performance of machine learning approaches Random Forest and a baseline logistic regression model employed in a prior BPSD prediction study [12] with our preprocessing approaches. These results are aggregated across all personalized models, representing the average performance. The experimental results reveal that the conventional approaches cannot achieve good performance without our proposed framework. To explore the expandability and flexibility of our proposed framework, we alternated different backbone models (classifiers) to validate the performance. TABLE IV displays the results of our proposed framework with various backbones, including RF and LR. Our framework can obviously benefit the performance of different backbone models on every metric. For example, the sensitivity of the baseline model LR has increased from 61.5% to 69.0%, relatively growing 12.2%, while the specificity keeps the same performance; for the RF model, the AUC has relatively increased 5.5%.

Notably, predicting BPSD within a short and specific timeframe (e.g., the next four hours) presents a challenge for conventional frameworks. Furthermore, the evaluation criteria of the personalized framework are more rigorous than traditional generalized frameworks. While the traditional approach primarily focuses on fine-tuning a single model, the personalized framework needs particular attention to numerous individual models. Consequently, our proposed framework is expertly tailored for fine-grained personalized BPSD prediction, making it a practical solution for real-world use.

### D. Ablation Study

TABLE V illustrates ablation studies assessing the effectiveness of different components of the proposed framework. The foundational model, the Raw-Data-based Generalized model (RDG), utilized the ERT algorithm for training. Various combinations of these components are investigated to distinguish the contributions of the Temporal Convolutional Network (TCN) and the Individual Representation-based Generalized (IRG) models. The RDG + TCN model exhibits a detectable performance degradation compared to the fundamental RDG model, yielding it less optimal. In contrast, the RDG + IRG model significantly enhances several metrics, underlining the value of incorporating individual representations in the generalized model. Remarkably, the amalgamation of RDG + IRG + TCN outperforms the other models and combinations, achieving superior results in metrics, including AUC, Sensitivity, Specificity, Accuracy, Precision, and F1-Score, highlighting the effect of integrating temporal contextual information and individualized representations with raw data-based modeling. The final combination generated a robust model that balances various classification metrics, thus providing a promising avenue for advancing state-of-the-art methodologies in the BPSD prediction research.

## V. Discussion and Conclusion

In this pioneering study, we have presented a personalized fine-grained BPSD prediction framework that utilizes physiological signals captured from smart wearable devices, marking a monumental advancement in dementia care and geriatric medicine. This study demonstrates the integration of raw data-based modeling, temporal contextual information, and tailored representations to accurately predict BPSD occurrences and categorize the fine-grained types of BPSD, including Hyperactivity, Psychosis, and Physical Behavioral symptoms.

A detailed comparative analysis between our personalized and conventional generalized methods unveils the significant breakthroughs and substantial advancements in this study. The personalized method has showcased improvements across all crucial performance metrics, emphasizing the efficacy of our approach in enhancing BPSD predictions. Our personalized model achieved a 16.0% in AUC, delineating a stark enhancement compared to the generalized model. The sensitivity and specificity improved significantly by 57% and 32.6%, respectively. The precision rate and F1-score also increased, demonstrating the impact of our method in this field. These experimental results revealed the potential for proactive interventions in dementia patients within a four-hour window, leading to improved outcomes and quality of life.

The future research will focus on expanding and diversifying datasets as well as optimizing models to achieve unprecedented levels of predictive accuracy. The ultimate goal is to ensure our models can be applied to a wide range of real-world scenarios. The innovative insights we gain from this research will revolutionize dementia care, leading to the development of personalized healthcare in geriatric medicine, and opening up new avenues for exploration in this field.

## Acknowledgment

This study was supported by National Health Research Institutes under grant no. NHRI-11A1-CG-CO-06-2225-1 and NHRI-12A1-CG-CO-06-2225-1.